# Humidity Sensing Properties of Different Atomic Layers of Graphene on SiO$_2$/Si Substrate

Qiang Gao[1,2], Hongliang Ma[1,2,3], Chang He[1,2], Xiaojing Wang[4]*, Jie Ding[3], Wendong Zhang[5,6]*, and Xuge Fan[1,2,3,7]*

[1] Advanced Research Institute for Multidisciplinary Science, Beijing Institute of Technology, Beijing 100081, China.

[2] Center for Interdisciplinary Science of Optical Quantum and NEMS Integration, Beijing Institute of Technology, 100081 Beijing, China.

[3] School of Integrated Circuits and Electronics, Beijing Institute of Technology, Beijing 100081, China.

[4] Advanced Interdisciplinary Technology Research Center, National Innovation Institute of Defense Technology, 100071, Beijing, China

[5] State Key Laboratory of Dynamic Measurement Technology, North University of China, Taiyuan 030051, China.

[6] National Key Laboratory for Electronic Measurement Technology, School of Instrument and Electronics, North University of China, Taiyuan 030051, China.

[7] Beijing Institute of Technology, Zhuhai, Beijing Institute of Technology, 519088 Zhuhai, China.

*E-mail: xgfan@bit.edu.cn, wang_xiaojing90@163.com, wdzhang@nuc.edu.cn

**ABSTRACT:** Graphene has the great potential to be used for humidity sensing due to ultrahigh surface area and conductivity. However, the impact of different atomic layers of graphene on $SiO_2$/Si substrate on the humidity sensing have not been studied yet. In this paper, we fabricated three types of humidity sensors on $SiO_2$/Si substrate based on one to three atomic layers of graphene, in which the sensing areas of graphene are 75 μm × 72 μm and 45 μm × 72 μm, respectively. We studied the impact of both the number of atomic layers of graphene and the sensing areas of graphene on the responsivity and response/recovery time of the prepared graphene-based humidity sensors. We found the relative resistance change of the prepared devices decreased with the increase of number of atomic layers of graphene under the same change of relative humidity. Further, devices based on tri-layer graphene showed the fastest response/recovery time while devices based on double-layer graphene showed the slowest response/recovery time. Finally, we chose the devices based on double-layer graphene that have relatively good responsivity and stability for application in respiration monitoring and contact-free finger monitoring.

**KEYWORDS:** graphene, humidity sensor, responsivity, response time, recovery time, respiration monitoring

## 1.INTRODUCTION

It is essential to monitor and control the humidity in numerous fields, such as healthcare,[1] environmental monitoring[2] and industrial production.[3] Humidity sensors are divided into resistive-type,[4] impedance-type,[5] capacitive-type,[6] surface acoustic wave (SAW)-type[7] etc., according to the working mechanism of the device. Nevertheless, humidity sensors made from conventional materials such as metal oxide,[8,9] metal/polymer composites[10] and ceramics,[11] generally suffer from high power consumption,[12] slow response/recovery time[13] and are not easily be integrated.[14] In addition, many advanced fabrication techniques (e.g., laser induction,[15,16] self-assembly,[17] etc.) combined with novel moisture-sensitive materials (e.g., black phosphorus (BP),[18] MXene,[19,20] etc.) have also been proposed for the fabrication of high-performance humidity sensors. Unfortunately, the above processing methods and materials are generally limited by complex processes and high costs. Hence, miniaturization, simplicity and low cost of sensors are highly required in the development of high-performance humidity sensors.

As one of the most widely used two-dimensional materials, graphene has been widely studied by researchers since its appearance in 2004. Graphene exhibits excellent electrical conductivity, high surface area ratio, better stability and are well compatible with advanced processing technique.[21–23] For example, Smith et al. experimentally proved that the much higher sensitivity of graphene to humid air than the main elements in air (dry Ar (argon), dry $N_2$ (nitrogen), dry $O_2$ (oxygen)), and prepared a humidity sensor based on monolayer CVD graphene on a $SiO_2$/Si substrate.[24] Yet, the impact of the number of atomic layers of graphene on graphene-based humidity sensors has not been studied. In addition, Wang et al. presented a BP-based humidity sensor which realizes the miniaturization of devices. Whereas, a 6 nm-thick $Al_2O_3$ layer needs to be deposited by ALD (atomic layer deposition) to protect BP from being oxidized, which complicates the processing of the sensor.[25] Further, Ma et al. presented a bacterial cellulose-based flexible humidity sensor with high sensitivity within the relative humidity ranging from 36.4% RH to 93% RH, which is difficult to be integrated due to material characteristics.[26] Besides, there are some reports about graphene-based humidity sensors such as the double-layer graphene-based humidity sensor,[27] the fluorinated graphene-based humidity sensor,[28] the wrinkled graphene-based humidity sensor[29] and the humidity sensor based on plasma modified graphene[30] etc., which provide the basis for the high-performance and miniaturization of graphene-based sensors. Meanwhile, it has been demonstrated that the change of humidity level has insignificant impact on contact resistance in graphene devices.[31] Although there are few reports about the impact of the number of atomic layers of graphene on the heat transport properties, charge density, electronic properties and optoelectronic properties of graphene-based materials and devices,[32–35] the impact of different atomic layers of graphene on $SiO_2$/Si substrate on the performance of graphene-based humidity sensors have not been studied. It would be meaningful to systematically study the impact of the number of atomic layers of graphene with different sensing areas on the performance of graphene-based humidity sensors, which would further pave the way of the development of humidity sensors based on graphene films and its derivatives.

Herein, we combined micro-nano fabrication technique with graphene-transfer process to manufacture humidity sensors on $SiO_2$/Si substrate by using monolayer graphene, double-layer graphene and tri-layer graphene respectively. We studied the impact of different atomic layers of graphene on the performance of the prepared humidity sensors such as sensitivity, response/recovery time. Moreover, the impact of the sensing areas (75 μm × 72 μm and 45 μm × 72 μm) of different atomic layers of graphene on the performance of prepared graphene-based humidity sensors was also experimentally studied. Finally, according to the characteristics of different atomic layers of graphene, we chose double-layer graphene-based devices to demonstrate their potential applications of respiration monitoring and contact-free finger monitoring for information interchange.

## 2. EXPERIMENAIL SECTION

### 2.1 Materials

CVD-grown monolayer graphene on copper foil was obtain from 6 Carbon Technology (Shenzhen, China). Poly (Bisphenol A carbonate) was produced by SIGMA-ALDRICH (USA) and Trichloromethane solution was obtained by TGREAG (Beijing, China). Iron chloride hexahydrate was purchased from RHAWN (Shanghai, China).

### 2.2 Device Substrate Manufacturing

Devices were fabricated from silicon substrates that is thermally oxidized to grow a 1.4 μm thick $SiO_2$ layer (Figure 1a). Four electrode contacts were fabricated by embedding the electrode metals (Ti/Au) into the $SiO_2$ layer. To be specific, a photoresist (PR) layer was spin-coated on the $SiO_2$ surface and patterned to define the metal electrode contacts. The 1.4 μm thick $SiO_2$ layer was patterned by etching 300 nm deep cavities using reactive ion etching (RIE) (Figure 1b). Next, a 50 nm titanium layer and a 270 nm gold layer were deposited inside the cavity by e-beam metal evaporation (Figure 1c) and ultimately about 20 nm electrodes were higher than $SiO_2$ surface.

**2.3 Graphene Transfer and Post-processing**

For the transfer of monolayer graphene, PC (Propylene carbonate 0.8575%wt) solution was firstly spin-coated on CVD-grown monolayer graphene on copper foil at 1000 rpm for 3 s and 3000 rpm for 45 s, and was baked at 90°C for 10 minutes. Carbon residues on the backside of the copper foil were removed using $O_2$ plasma etching. Then, PC/graphene was then placed onto the surface of a $FeCl_3$ solution (6.4% wt), resulting in wet etching of the copper. Afterwards, the PC/graphene stack without copper, was transferred from the $FeCl_3$ solution onto the surface of deionized (DI) water, then to diluted HCl solution and back to DI water for cleaning. These cleaning steps were performed to remove iron (III) residues and chloride residues, respectively. The PC/graphene stack floating on the DI water was transferred to the surface of the pre-processed $SiO_2$/Si substrate (Figure. 1d). The chip was then baked for 10 min at 45°C to dry it. Next, the chip was placed into chloroform to dissolve the remaining PC from the graphene surface, followed by annealing at 300°C for 2 h to further dissolve the PC residues (Figure 1e). Double-layer and tri-layer graphene were obtained by stacking monolayer graphene samples based on the same processes previously described for copper removal from the monolayer graphene, and were transferred to the surface of the pre-processed $SiO_2$/Si substrate. Finally, the graphene was selectively patterned by the optical lithography (Tuotuo Technology (Suzhou) Co. Ltd.) and oxygen plasma etching with low power (Figure 1f).[36] Next, the graphene chips were glued in a ceramic package and devices were wire bonded (Figure 1h).

## 2.4. Characterization

The prepared devices were characterized by scanning electron microscopy (SEM) Apreo C (Thermo Fisher Scientific, USA). SEM images of two wire bonded humidity sensors based on tri-layer graphene were displayed in Figure 1g and 1i, respectively. More SEM images of devices based on monolayer graphene, double-layer graphene and tri-layer graphene with different sensing areas of graphene can be seen in Figure 2a-f. Raman spectrums of monolayer, double-layer and tri-layer graphene on $SiO_2$/Si substrate with typical G peak and 2D peak as well as quite weak D peak were obtained, indicating relatively good quality of these graphene samples after they were transferred from copper substrate to $SiO_2$/Si substrate (SI Figure S1).

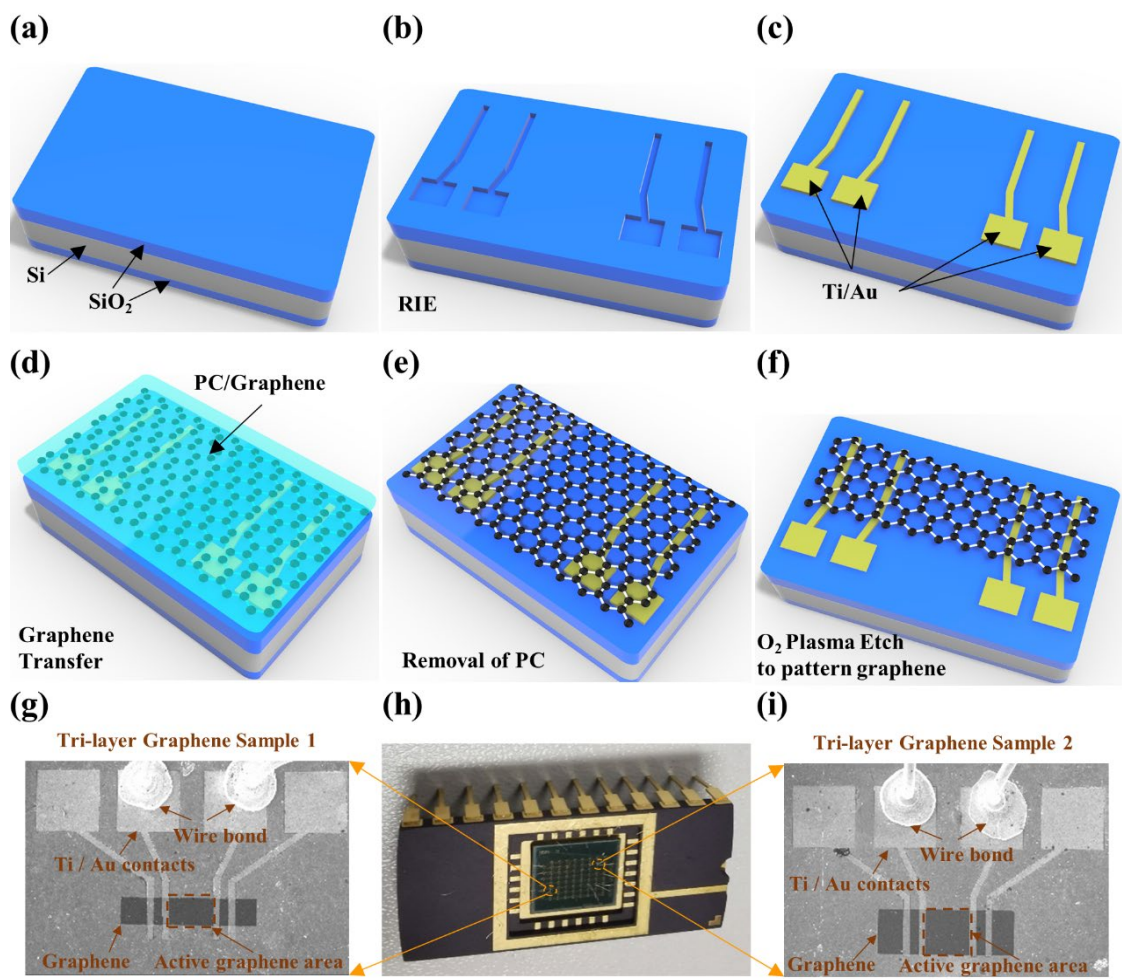

**Figure 1.** (a-f) Schematic of fabrication process of humidity sensors of different atomic layers of graphene on SiO$_2$/Si substrate. (g, i) SEM images of two wire bonded devices based on tri-layer graphene from a packaged chip. (h) Photograph of a packaged chip.

The experimental setup for humidity sensing experiments consists of a power supply, a data acquisition system, and a vacuum chamber which is connected via pipes to an Argon (Ar) tank, a vacuum pump system and a humidifier (SI Figure S2a). The ceramic package with the patterned monolayer/double-layer/tri-layer graphene devices were placed on a connector board, which was loaded in the vacuum chamber. A commercial humidity sensor (HIH-4000, Honeywell International Inc.) was put alongside the graphene devices in the vacuum chamber (SI Figure S2b). Electrical output of the graphene devices was acquired by a SMU (2450 SourceMeter, Keithley). The commercial humidity sensor was connected to DAQ card (Art Technology Data Acquisition Device) for simultaneous data acquisition.

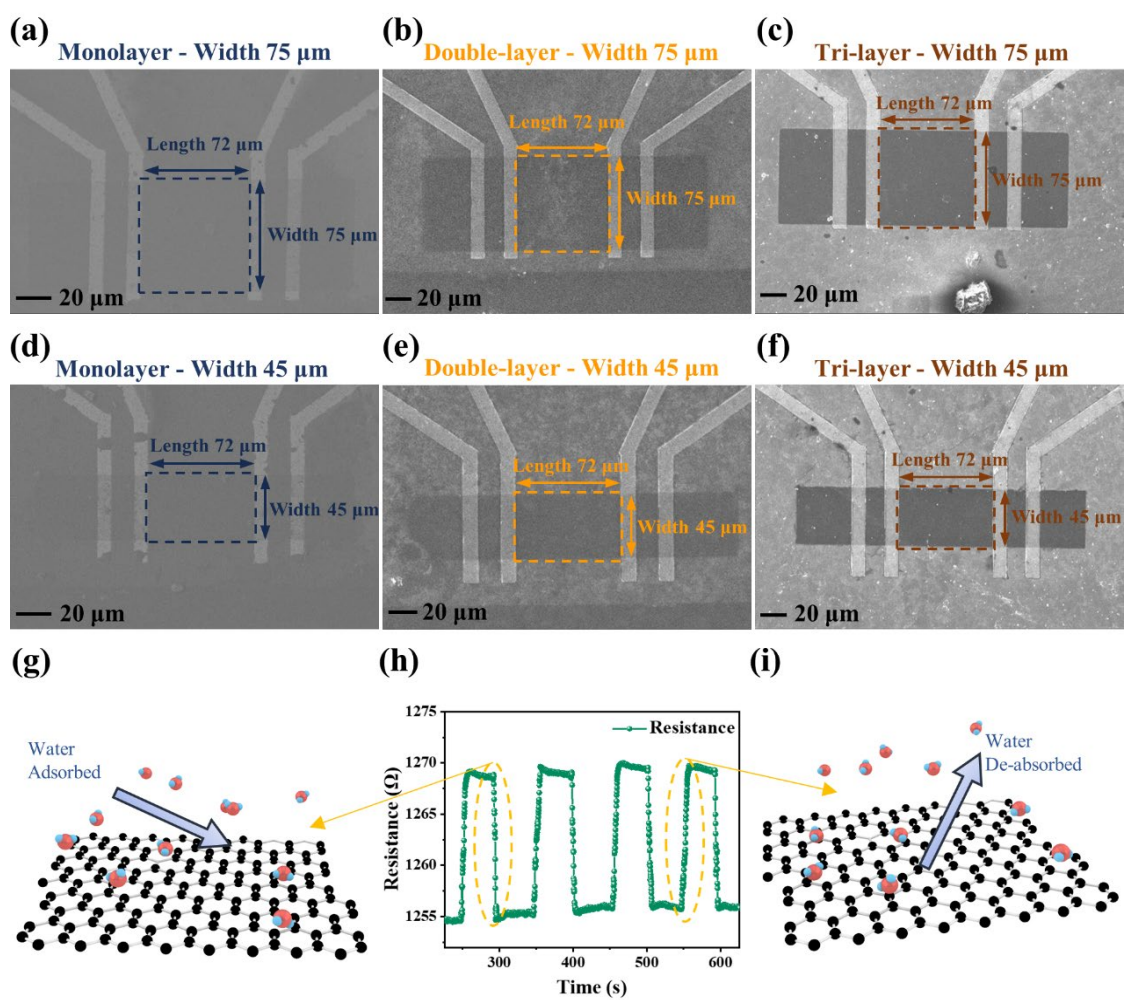

**Figure 2.** (a-f) SEM images of monolayer/double-layer/tri-layer graphene-based humidity sensors on $SiO_2$/Si substrate with the sensing areas of 75 μm × 72 μm and 45 μm × 72 μm. (h) The resistance response of a tri-layer graphene device during the cyclic humidity change from 92% RH to 10% RH. (g, i) Schematic diagrams of water molecules absorbing on and desorbing from graphene surface.

In the humidity sensing experiment, air was firstly evacuated from the vacuum chamber and dry Ar (argon) was slightly pumped into the chamber to rapidly decrease the relative humidity from 42% RH to approximate 10% RH. When the relative humidity in the vacuum chamber reached 10% RH, the argon was completely pumped out from the vacuum chamber, resulting in the decreased pressure in the vacuum chamber. Then, the air passed through the humidifier and entered into the vacuum chamber, and thereby the humidity level of the vacuum chamber restored to 42% RH. Although the concentration of $N_2$ (nitrogen), Ar and $O_2$ (oxygen) changed during the humidity measurement process, compared to the humidity, the concentration change of $N_2$, Ar and $O_2$ had a negligible impact on the resistance of graphene-based devices.[24] With the change of relative humidity in the vacuum chamber, the electrical resistance of graphene devices changed accordingly (Figure 2h). When the relative humidity in the chamber was increased, more water molecules adsorbed on graphene surface causing substantial charge transferring from the $SiO_2$ substrate with $Q^0_3$ defect to graphene (Figure 2g).[24] Likewise, when the relative humidity was decreased, water molecules desorbed from $SiO_2$ substrate (Figure 2i). A high precision SMU (2450 Sourcemeter, Keythley) was used to capture graphene devices' real-time resistance. To avoid possible destruction of graphene devices, a low current of 70 μA was used for measuring the resistance of graphene devices. The output signal of the commercial humidity sensor (HIH-4000, Honeywell International Inc.) was collected by a high precision data acquisition card (Art Technology Data Acquisition Device) to monitor the humidity level in the vacuum chamber. In our experiments, we respectively prepared humidity sensors based on monolayer/double-layer/tri-layer graphene with two graphene sensing areas of 75 μm × 72 μm and 45 μm × 72 μm. Each of them was used for humidity sensing experiments to study the impact of the number of atomic layers of graphene and sensing areas of graphene on the responsivity and response/recovery time.

## 3. RESULTS AND DISCUSSION

### 3.1 The Impact of the Number of Atomic Layers of Graphene and Sensing Areas of Graphene on Devices' Responsivity

Figure 2a-f showed SEM images of monolayer/double-layer/tri-layer graphene devices with different sensing areas (75 μm × 72 μm and 45 μm × 72 μm), in which the dashed line frame marked the dimensions of the sensing unit of each device. The responsivity of graphene-based humidity sensors was evaluated by measuring its resistance variation under different RH conditions (10% RH to 42% RH) at room temperature. The responsivity of the prepared humidity sensor can be calculated by responsivity = $\Delta R/R_0 \times 100\%$, where $\Delta R = R - R_0$, $R_0$ is the measured initial resistance of the device at 42% RH, and R is the actual resistance under certain relative humidity level. As shown in Figure 3a-f, the electrical resistance response of monolayer/double-layer/tri-layer graphene-based humidity sensors with sensing areas of 75 μm × 72 μm and 45 μm × 72 μm was measured as the relative humidity was decreased from 42% RH to 10% RH for three consecutive cycles. The resistance of such all devices well followed with the change of the relative humidity and was increased with the decrease

of the relative humidity, featuring excellent stability and repeatability. Likewise, as the relative humidity decreased from 22% RH to 10% RH, the responsivity of graphene-based devices decreased with the increase of the number of atomic layers of graphene under the conditions of the same sensing area of graphene film (SI Figure S3). It should be noted that double-layer and tri-layer graphene were obtained by stacking CVD-grown monolayer graphene, and thereby were not AB stacked, which enabled double-layer graphene and tri-layer graphene to have the identical trend of resistance change with monolayer graphene due to the weak coupling between each graphene atomic layer, as the relative humidity was changed.[35,37]

Figure 3g compared the responsivity of the prepared devices with different number of atomic layers of graphene and different sensing areas of graphene that were shown in Figure 3a-f. The responsivity of graphene-based devices decreased with the increase of the number of atomic layers of graphene under the conditions of the same sensing area of graphene film and the same decrease of relative humidity from 42% RH to 10% RH. For instance, for the graphene sensing area of 75 μm × 72 μm, the relative resistance change of monolayer graphene-based device is over 5 times larger than that of tri-layer graphene-based device. According to the relevant literatures,[24,38] an impurity band exists on the $SiO_2$ surface with the $Q^0_3$ defect and can shift due to the electro-static dipole moment of the water molecules, which results in an effective doping and thereby increases the conductivity of the graphene layer. With the increase of the number of graphene atomic layers, the interaction of the $SiO_2$ substrate and the top graphene atomic layer will weaken, which decreases the doping effect induced by water molecules and thereby decreases the responsivity of the humidity sensors. Therefore, the monolayer graphene-based devices show the highest responsivity, while tri-layer graphene-based device show the lowest responsivity.

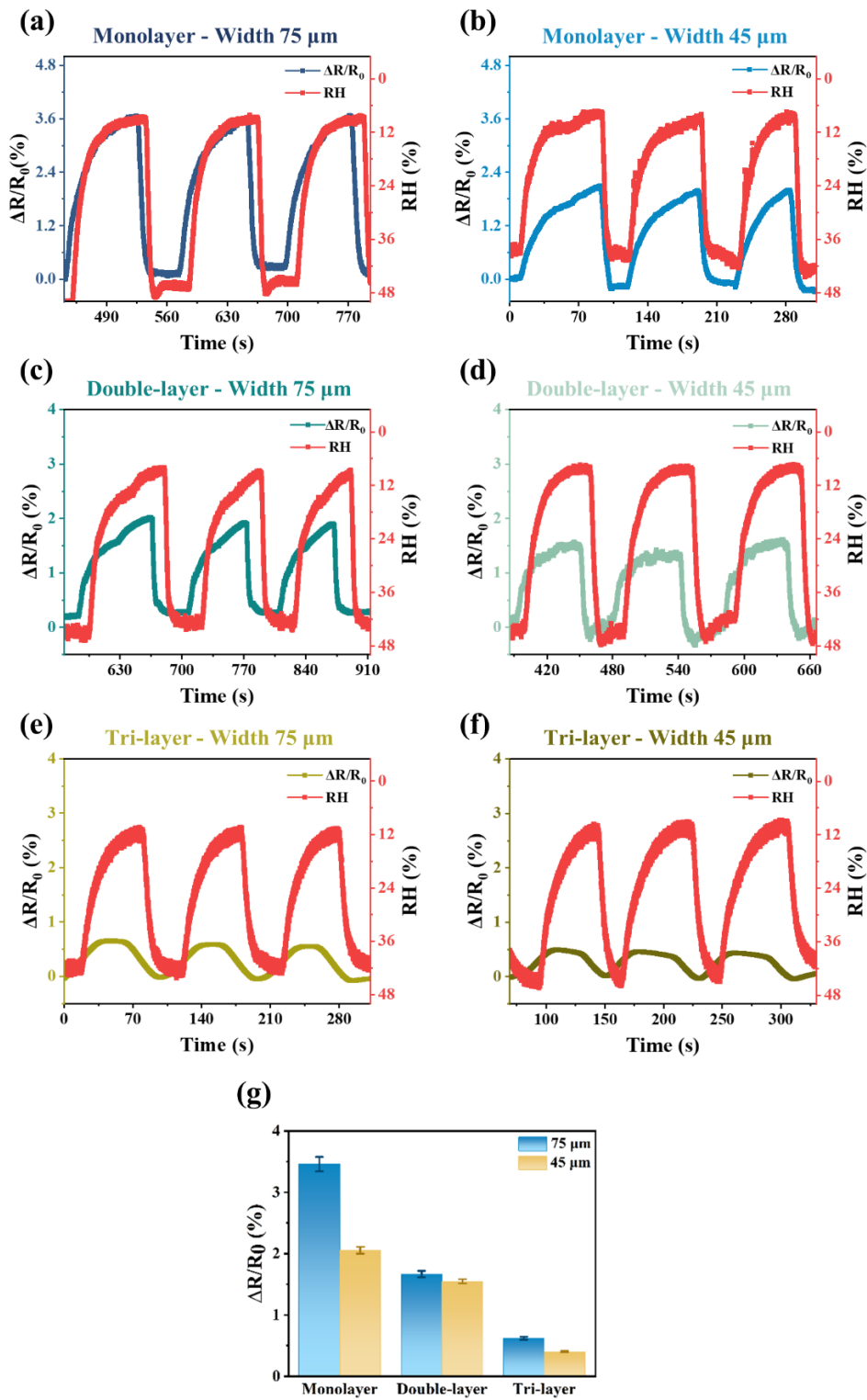

**Figure 3.** (a-f) Resistance response of graphene-based devices in Figure 2 (a-f) with different number of atomic layers of graphene and different sensing areas of graphene as the relative humidity was changed between 42% RH to 10% RH for three consecutive cycles. (g) Comparison of relative resistance change of graphene-based devices in (a-f) as the relative humidity decreased from 42% RH to 10% RH in three consecutive humidity measurement cycles.

Figure 3g also shows that the responsivity of graphene-based devices decreased with the decrease of sensing area of graphene under the condition of the same number of the atomic layers of graphene. For instance, for the same length (72 μm length) of the graphene sensing film, wider graphene films (75 μm width) have the higher responsivity than narrow graphene films (45 μm width). This can be attributed to the more adsorption sites of water molecules of large sensing area of graphene compared to small sensing area of graphene.

**3.2 The Impact of Graphene Layers on Response/Recovery Time**

As the important indicators of the humidity sensor, response and recovery time are defined as the time required to achieve ~90% of resistance change during the adsorption and desorption process.[39] Figure 4a-f illustrate the response/recovery time of monolayer/double-layer/tri-layer graphene-based devices with the graphene sensing area of 75 μm × 72 μm as the relative humidity changed between 7% RH to 92% RH.

As shown in Figure 4a and b, the response/recovery time of two monolayer graphene-based devices are approximately 8.64 s/9.45 s and 6.66 s/12.59 s respectively. For two double-layer graphene-based devices, their response/recovery time are approximately 17.21 s/34.06 s and 12.26 s/21.23 s respectively (Figure 4c and d). For two tri-layer graphene-based devices, their response/recovery time are approximately 4.96 s/9.89 s and 5.09 s/8.67 s respectively (Figure 4e and f). Figure 4g and h compared the response/recovery time of such six devices with different number of atomic layers of graphene. The relatively fast response and recovery time of the monolayer graphene-based devices benefit from the single atomic layer thickness of monolayer graphene. However, for two double-layer graphene-based devices, the bottom atomic layer of graphene was affected by the $SiO_2$ substrate,[40] with better hydrophilia, and thereby can capture water molecules slowly from the defects of the top graphene atomic layer.[41] This would lengthen the process of water molecules that are adsorbed on the graphene surface and thereby increase the response time. Meanwhile, as the relative humidity decreased, the bottom atomic layer of double-layer graphene will prevent water molecules from escaping the gap between the double graphene out, leading to relatively slow recovery time. By contrast, tri-layer graphene-based devices show the fastest response and recovery time among all devices. This is because the thickness of tri-layer graphene is further increased, resulting in the decreased impact of $SiO_2$ substrate on the humidity sensing properties of graphene-based devices. In addition, the characteristics of tri-layer graphene is close to graphite,[40] resulting in the decreased humidity responsivity and fast response/recovery time.

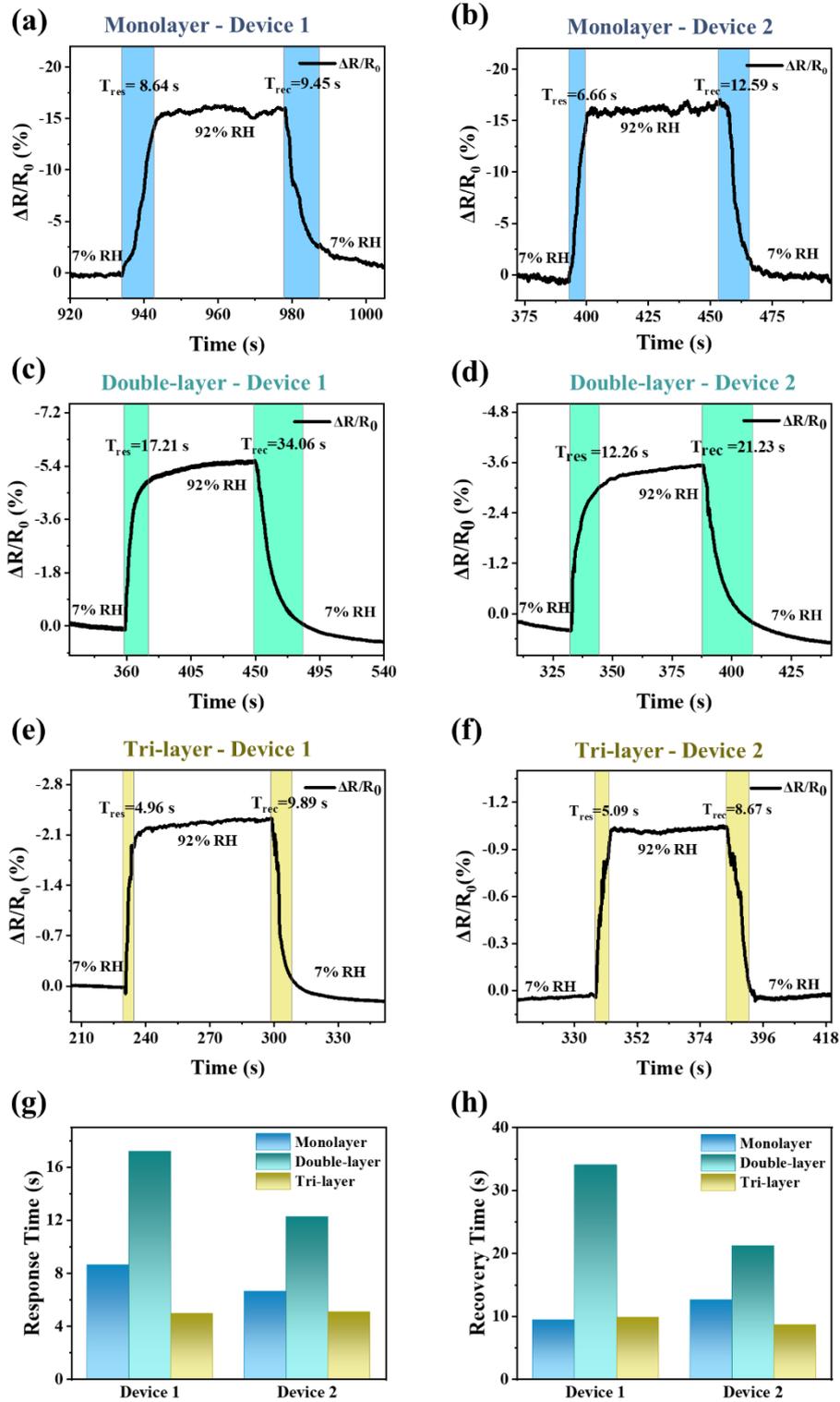

**Figure 4.** (a-f) The response and recovery time of six devices based on monolayer graphene (a, b), double layer graphene (c, d) and tri-layer graphene (e, f) as the relative humidity was changed between 7% RH and 92% RH. (g, h) The comparison of response/recovery time of six graphene devices with different number of atomic layers of graphene in (a-f).

**3.3 The Application of the Double-layer Graphene-based Humidity Sensor**

In comparison to double-layer graphene-based devices, monolayer graphene-based devices were generally limited by the stability (SI Figure S4), and tri-layer graphene-based devices were generally limited by the responsivity. Taking the relatively good responsivity and stability of double-layer graphene-based devices into consideration, double-layer graphene-based devices were used for human breathing monitoring and contact-free finger monitoring. Figure 5a-c illustrate the electrical response of double-layer graphene-based devices when subjected to fast, normal and deep breathing of people. As the frequency of the breathing slows down from fast breathing, normal breathing to deep breathing, the response periods of the device were about 0.74 s, 1.5 s and 2.1 s respectively, and the relative resistance change of the device increased. What's more, the device based on double-layer graphene was also used for contact-free finger monitoring. The finger was put on top of the device with 5 mm height gap for 2 s and 4 s, respectively, and the resistance of the device decreased correspondingly. As the resistance of the device recovered to the original value, then the finger was again put on top of the device with 5 mm height gap for 2 s and 4 s, respectively. And the cycle repeated. As shown in Figure 5d, the electrical response of the double-layer graphene-based device to the finger that was put on top of the device for 2 s displayed the shorter response period and smaller relative resistance change, compared to the situation of 4

s. These results demonstrate the potential application of prepared graphene-based humidity sensors for human respiration monitoring and the contact-free information change. Due to the low temperature coefficient of graphene film[42] and the short time for finger action, the impact of temperature on graphene-based humidity sensors can be ignored, taking into consideration the fact that the finger was not directly in touch with the surface of graphene.

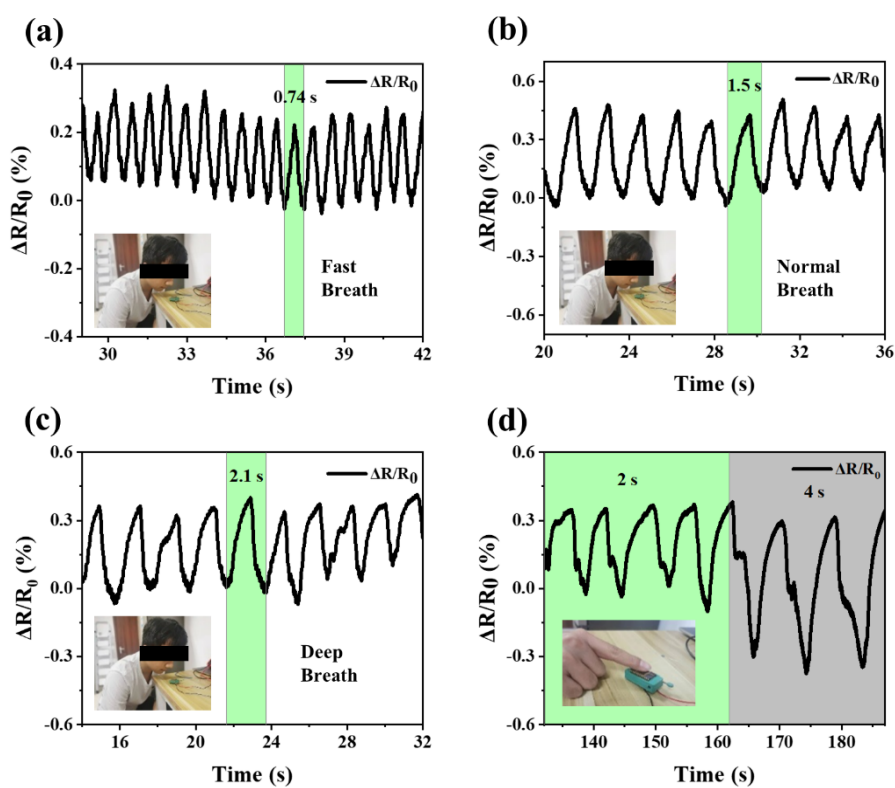

**Figure 5.** (a-c) The electrical response of the double-layer graphene-based device to fast, normal and deep breathing of people respectively. (d) The electrical response of the double-layer graphene-based device to the people's finger that was put on top of device with 5 mm height for 2 s and 4 s, respectively.

## 4. CONCLUSIONS

In this work, we have successfully developed graphene-based humidity sensors with different number of atomic layers of graphene. The impact of the number of atomic layers of graphene on the responsivity and response/recovery time of prepared humidity sensors is experimentally studied. The monolayer graphene-based devices show the highest responsivity while the tri-layer graphene-based devices show the lowest responsivity. This can be attributed to the decreased impact of the $SiO_2$ substrate on the humidity sensing performance of top atomic layer of graphene in tri-layer graphene-based device. Further, devices based on monolayer graphene and tri-layer graphene show shorter response/recovery time compared to devices based on double-layer graphene. The double-layer graphene-based humidity sensors that have relatively good responsivity and stability were applied for human respiration monitoring and contact-free information exchange. These findings would contribute to the understanding of humidity sensing properties of graphene-based humidity sensors on $SiO_2/Si$ substrates.

## ASSOCIATED CONTENT

**Supporting Information**

The Supporting Information is available free of charge on the ACS Publications website at http:// pubs.acs.org.

Raman characterization of monolayer/double-layer/tri-layer graphene on $SiO_2/Si$ substrate (Figure S1); Experimental setup and measurement circuit (Figure S2); Resistance response of graphene-based devices in Figure 2 (a-f) with different number of atomic layers of graphene and different sensing areas of graphene (Figure S3); The description about the stability of monolayer, double-layer and tri-layer based humidity sensors (Figure S4).

## AUTHOR INFORMATION

**Corresponding Authors**

*E-mail: xgfan@bit.edu.cn, wang_xiaojing90@163.com, wdzhang@nuc.edu.cn

**Notes**

The authors have no competing financial interest.

## ACKNOWLEDGMENTS

This work was supported by Beijing Natural Science Foundation (4232076), National Natural Science Foundation of China (62171037 and 62088101), 173 Technical Field Fund (2023-JCJQ-JJ-0971), National Key Research and Development Program of China (2022YFB3204600), Beijing Institute of Technology Science and Technology Innovation Plan (2022CX11019), National Science Fund for Excellent Young Scholars (Overseas), Beijing Institute of Technology Teli Young Fellow Program (2021TLQT012). The SEM characterization was performed at the Analysis & Testing Center of Beijing Institute of Technology.

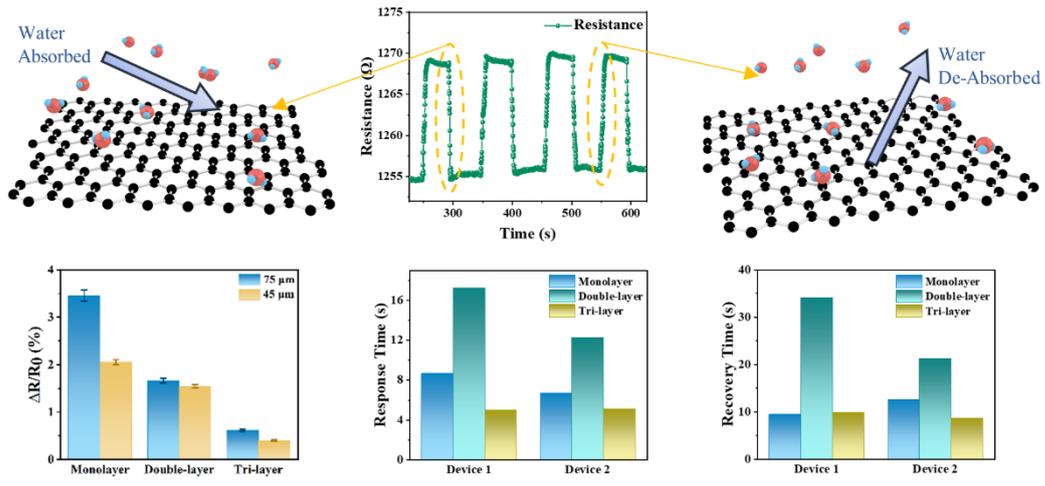